\documentclass[twocolumn,showpacs,preprintnumbers,amsmath,amssymb]{revtex4}

\usepackage{graphicx}
\usepackage{dcolumn}
\usepackage{bm}

\begin{document}

\preprint{APS/PRL}

\title{Dome-shaped magnetic phase diagram of thermoelectric layered cobaltites}

\author{J. Sugiyama$^1$}
 \email{e0589@mosk.tytlabs.co.jp}
\author{J. H. Brewer$^2$}
\author{E. J. Ansaldo$^3$}
\author{H. Itahara$^1$}%
\author{T. Tani$^1$}%
\author{M. Mikami$^4$}%
 \altaffiliation[Present address: ]{National Institute of Advanced Industrial Science and technology, Ikeda, Osaka 563-8577, Japan.}
\author{Y. Mori$^4$}
\author{T. Sasaki$^4$}
\author{S. H{\' e}bert,$^5$}
\author{A. Maignan$^5$}
\affiliation{%
$^1$Toyota Central Research and Development Labs. Inc., 
 Nagakute, Aichi 480-1192, Japan}%

\affiliation{$^2$TRIUMF, CIAR and 
Department of Physics and Astronomy, University of British Columbia, 
Vancouver, BC, V6T 1Z1 Canada 
}%

\affiliation{$^3$TRIUMF, 4004 Wesbrook Mall, Vancouver, BC, V6T 2A3 Canada 
}%

\affiliation{$^4$Department of Electrical Engineering, Osaka University, 
Suita, Osaka 563-8577, Japan
}%

\affiliation{$^5$Laboratoire CRISMAT, CNRS/ISMRA/University of Caen, 
6 bd du Mar{\' e}chal Juin, 14050 Caen Cedex, France
}%

\date{\today}

\begin{abstract}
Using muon spin spectroscopy 
we have found that, 
for both Na$_x$CoO$_2$ 
(0.6 $\leq x \leq$ 0.9) 
and 3- and 4-layer cobaltites, 
a common low temperature magnetic state 
(which in some cases is manifest as an incommensurate spin density wave) 
forms in the CoO$_2$ planes. 
Here we summarize those results and report 
a dome-shaped relation between the transition temperature 
into the low-$T$ magnetic state and the composition $x$ 
for Na$_x$CoO$_2$ and/or the high-temperature asymptotic limit of 
thermopower in the more complex 3- and 4-layer cobaltites. 
This behavior is explained using the Hubbard model 
on two-dimensional triangular lattice 
in the CoO$_2$ plane.
\end{abstract}

\pacs{76.75.+i, 75.30.Fv, 72.15.Jf, 75.30.Kz}%
\keywords{Thermoelectric layered cobaltites, magnetism, 
 muon spin rotation, incommendurate spin density waves}

\maketitle

Although the widespread current interest in the layered cobaltites 
\cite{NCO_sc_1, NCO_sc_2}
was originally due mainly to their unique combination of 
high thermopower $S$ with metallic transport properties,
\cite{NCO_1,NCO_2,NCO_3,CCO_1,CCO_2, PbCSCO_1,TlSCO_1,HgSCCO_1,BSCO_1,CCCO_1}
which makes them one of the most promising systems for power applications, 
we have shown that they also display interesting and 
complex magnetic orderings, directly correlated 
with the enhanced thermopower.
The richness of behavior of the layered cobaltites is 
due to their intrinsic structure, 
namely: electrically active triangular plane layers of CoO$_2$, 
which are separated by a variety of intermediate structures; 
their relatively strong electronic correlations; 
and the fact that the structures between the CoO$_2$ layers 
can be modified in a variety of ways to vary their dimensionality, 
ionic states, carrier doping in the CoO$_2$ planes and 
the relevant interaction strengths.

In order to elucidate the magnetism in the CoO$_2$ planes and 
the mechanism of the good thermoelectric properties, 
we have carried out positive muon spin rotation and relaxation 
($\mu^+$SR) experiments on the layered cobaltites. 
As a result, we found the transition from 
a high-temperature paramagnetic  
to a low-temperature commensurate or incommensurate 
spin density wave ({\sf C- or IC-SDW}) state 
for [Ca$_2$CoO$_3$]$_{0.62}^{\rm RS}$[CoO$_2$] 
below $\sim$100~K,\cite{muSR_1,muSR_3} 
Na$_{0.75}$CoO$_2$ at 22~K \cite{muSR_2} and
[Ca$_2$Co$_{4/3}$Cu$_{2/3}$O$_4$]$_{0.62}^{\rm RS}$[CoO$_2$] 
below $\sim$200~K,\cite{muSR_4} 
where RS denotes the rocksalt-type subsystem. 
The common magnetic ordering is thus not always 
that of a classic frustrated {\sf AF} 
(triangular planar arrangement of spins) 
system but rather a variety of states including disordered {\sf AF} 
and (most common) an incommensurate spin density wave ({\sf IC-SDW}), n
arising in parallel with other effects in the triangular lattice 
such as effective mass enhancement and the enhanced $S$ 
in the Na$_x$CoO$_2$ case. 
The more complex cobaltites exhibit, in addition, 
other magnetic orderings and even higher $S$. 

Nevertheless, the transition temperature was scattered 
in the wide temperature range (22--200~K), 
and there was no clear relationship between $T_{\sf SDW}$ 
and structural properties and/or carrier concentration 
of the layered cobaltites. 
We back therefore to the basic system, Na$_x$CoO$_2$, 
to investigate the dependence of $T_{\sf SDW}$ on $x$ by $\mu^+$SR.
Also, for the multilayer systems, 
we clarify the magnitude of $T_{\sf SDW}$ 
as a function of the Co valence varied by the change 
in composition of the rocksalt-type subsystem.
\cite{PbCSCO_1,TlSCO_1,HgSCCO_1} 
For the Na$_x$CoO$_2$ case, 
we show that a model based on the Hubbard model 
for a triangular lattice of {\bf S}=1/2 Co ions with occupancy given by $x$ 
(correlated with band filling)\cite{2DTL_1,2DTL_2} is able to explain 
the dome shaped magnetic phase diagram shown below. 
Similar behavior is observed for the multilayer systems, 
where the increased two dimensionality due to the extra layers 
leads to higher $T_{\sf SDW}$ for the {\sf IC-SDW} state and 
is correlated with higher $S$.

Single-crystal platelets of Na$_x$CoO$_2$ were prepared 
by a flux method.\cite{NCO_4} 
Polycrystalline 3- and 4-layer cobaltites listed in Table~1 
were synthesized by a solid state reaction technique 
\cite{muSR_1,PbCSCO_1,TlSCO_1,HgSCCO_1} 
or a reactive templated grain growth technique.\cite{rtgg_1} 
The $\mu^+$SR experiments were performed on the {\bf M15} and {\bf M20} 
surface muon beam lines at TRIUMF. 
The experimental setup is described in elsewhere.\cite{SDW_1,SDW_2}

\begin{figure}
\includegraphics[width=8cm]{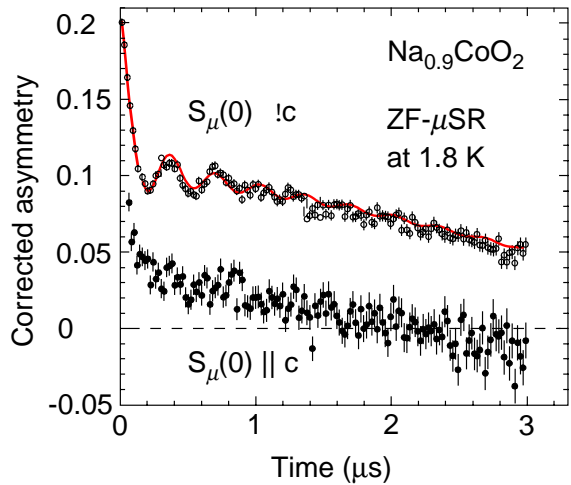}
\caption{\label{fig:ZF-muSR1} ZF-$\mu^+$SR time spectra of 
  single crystal platelets of Na$_{0.9}$CoO$_2$ at 1.8~K. 
  The configurations of the sample and the initial muon spin direction 
  $\vec{\bm S}_\mu(0)$ are 
  (top) $\vec{\bm S}_\mu(0) \perp \hat{\bm c}$ and 
  (bottom) $\vec{\bm S}_\mu(0) \parallel \hat{\bm c}$.  
  The bottom spectrum is offset below to be seen clearly.
}
\end{figure}
Figure~1 shows zero-field (ZF-)$\mu^+$SR time spectra at 1.8 K 
for single crystal platelets of Na$_{0.9}$CoO$_2$. 
The top spectrum was obtained with the initial $\mu^+$ spin direction 
$\vec{\bm S}_\mu(0)$ perpendicular to the $c$-axis 
and the bottom one with $\vec{\bm S}_\mu(0)$ parallel to $c$. 
A clear oscillation due to quasi-static internal fields is observed 
only for $\vec{\bm S}_\mu(0)$ perpendicular to ${\bm c}$. 
The muon signal is fitted best by a zeroth-order Bessel function 
of the first kind, $J_0(\omega_{\mu}t)$, 
which describes the evolution of the muon polarization 
in an {\sf IC-SDW} field distribution; 
\cite{SDW_1,SDW_2} 
we therefore conclude that Na$_{0.9}$CoO$_2$ undergoes 
a magnetic transition from a paramagnetic state to 
an {\sf IC-SDW} state at $T_{\sf SDW}$=19~K. 
The absence of a clear oscillation 
for the $\vec{\bm S}_\mu(0) \parallel {\bm c}$ case 
indicates that the internal field ${\bm H}_{\rm int}$ is 
roughly parallel to the $c$-axis.
Because of the strong anisotropy, the {\sf IC-SDW} is most likely to propagate 
in the $c$ plane ({\sl i.e.} in the CoO$_2$ plane), 
with oscillating moments directed along the $c$-axis. 

\begin{figure}
\includegraphics[width=8cm]{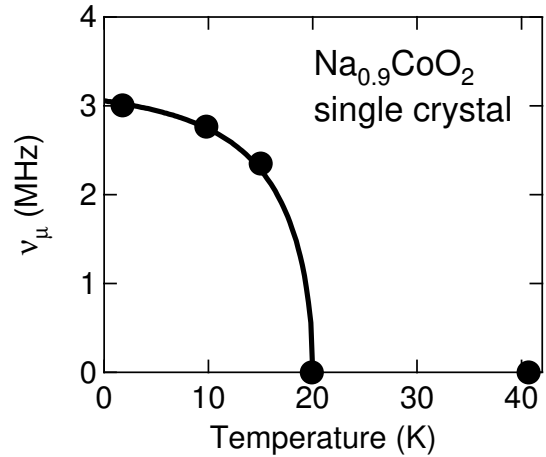}
\caption{\label{fig:ZF-muSR2} Temperature dependence of 
  the muon precession frequency 
  $\nu_{\mu}$ in Na$_{0.9}$CoO$_2$. 
  The solid line represents the temperature dependence 
  of the {\sf BCS} gap energy.
}
\end{figure}
Figure~2 shows the temperature dependence of the muon precession frequency
$\nu_{\mu}$ =$\omega_{\mu}/2\pi$ for the single crystal platelets of 
Na$_{0.9}$CoO$_2$. 
Here $\nu_{\mu}$ is the order parameter of the transition and 
its $T$ dependence is well described by the {\sf BCS} weak coupling 
expression for such order parameters, 
as expected for the {\sf IC-SDW} state.\cite{SDW_4}
\begin{figure}
\includegraphics[width=8cm]{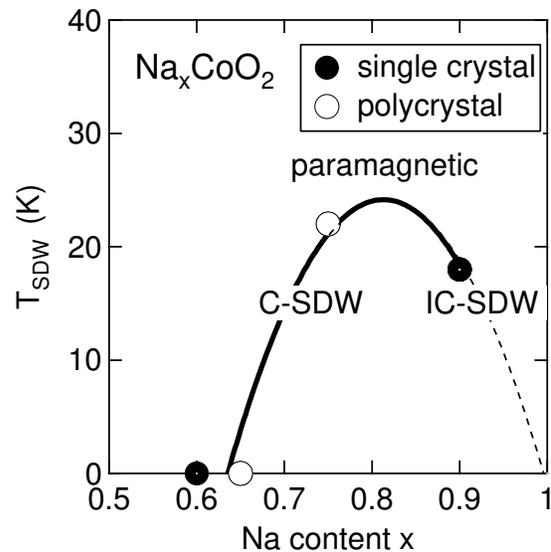}
\caption{\label{fig:ZF-muSR}  Phase diagram of Na$_x$CoO$_2$ 
  determined by the $\mu^+$SR experiments. 
  Solid and open circles represent 
  the present results on single crystals and polycrystalline samples, 
  respectively. 
  The point at $x$=1 is extrapolated from 
  the data on the related compound LiCoO$_2$.\cite{LCO_1}
}
\end{figure}
The magnetic phase diagram (Fig.~3) of Na$_x$CoO$_2$ can thus be 
sketched from the $\mu^+$SR results for polycrystalline samples 
with $x$=0.65 and 0.75 \cite{muSR_2} and the present $x$=0.6 and 0.9 crystals. 
Recent compositional and chemical titration analyses indicated that 
the oxygen deficiency $\delta$ in Na$_x$CoO$_{2-\delta}$ is 
negligibly small even for the $x$=0.9 sample.\cite{NCO_4} 
Hence, the average Co valence can be directly calculated from $x$. 
As $x$ increases from 0.6 ({\sl i.e.} the Co valence decreases from 3.4), 
the magnitude of $T_{\sf SDW}$ increases up to around $x$=0.8, 
then decreases with further increasing $x$. 
As a result, we obtain the dome-shaped relationship of Figure~3, 
between $T_{\sf SDW}$ and $x$, {\sl viz.} the Co valence for Na$_x$CoO$_2$.

The $x$=1 end member of Na$_x$CoO$_2$, that is, 
a fully occupied NaCoO$_2$ phase, 
cannot be prepared by conventional solid state reaction 
and/or flux techniques, 
whereas the related compound LiCoO$_2$ is easily obtained. 
The structure of LiCoO$_2$ is isomorphous with $\alpha$-NaFeO$_2$ ($R3m$), 
almost the same as that of Na$_x$CoO$_2$, 
and it is reported to be diamagnetic down to 4.2~K.\cite{LCO_1} 
This behavior is expected, because the Co$^{3+}$ ions in LiCoO$_2$ are 
in a low-spin state $t_{2g}^6$, as for Na$_x$CoO$_2$. 
Therefore, Na$_x$CoO$_2$ is also expected to lack magnetically ordered states. 

The occupancy of Co$^{4+}$ spins ({\bf S}=1/2) in 
the two-dimensional triangular lattice ({\sf 2DTL}) increases 
with decreasing $x$. 
Thus the other end member, Na$_0$CoO$_2$, 
would be a half filled {\sf 2DTL}. 
In other words, every lattice site is occupied by an {\bf S}=1/2 spin. 
The Hubbard model within a mean field approximation can be used 
for explaining the magnetism of such a system, 
with the Hamiltonian ${\cal H}$ given by
\cite{2DTL_1,2DTL_2}
\begin{eqnarray}
 {\cal H}&=&-t\sum_{<ij>\sigma}c_{i\sigma}^{\dagger}c_{j\sigma} + 
 U\sum_i n_{i\uparrow}n_{i\downarrow} ,
\label{eq:Hubbard}
\end{eqnarray}
where $c_{i\sigma}^{\dagger}(c_{j\sigma})$ 
creates (destroys) an electron with spin $\sigma$ on site $i$, 
$n_{i\sigma}=c_{i\sigma}^{\dagger}c_{i\sigma}$ 
is the number operator, 
$t$ is the nearest-neighbor hopping amplitude and 
$U$ is the Hubbard on-site repulsion.
The electron filling $n$ is defined as 
$n$=(1/2$N$)$\sum_i^N n_i$,
where $N$ is the total number of sites.  

At $T$=0 and $n$=0.5 ({\sl i.e.}, 
Na$_0$CoO$_2$), 
as $U$ increases from 0, the system is a paramagnetic metal 
up to $U/t$=3.97 due to geometrical frustration, 
then changes into a metal with a spiral {\sf IC-SDW}, 
and then at $U/t$=5.27 
a first-order metal-insulator transition occurs.\cite{2DTL_1} 
The lack of magnetic transitions for Na$_x$CoO$_2$ 
with $x$=0.6 and 0.65 suggests that $U/t\leq$3.97. 
This means that Na$_x$CoO$_2$ is unlikely to be 
a typical strongly correlated electron system, 
because $U\gg t$ for such a system.
The calculations\cite{2DTL_2} also predict that, as $n$ increases from 0, 
the magnitude of $U/t$ at the boundary between 
the paramagnetic and {\sf SDW} phases decreases, 
with increasing slope ($d(U/t)/dn$) up to $n$=0.75. 
Even for $U/t$=0, the {\sf SDW} phase is stable at $n$=0.75. 
The value of $U/t$ then increases with further increasing $n$, 
with decreasing slope. 
Therefore, the dome-shaped phase diagram of Fig.~3 
is qualitatively explained by the calculations, 
although the measured maximum of the dome is located around $x$=0.8 
({\sl i.e.} $n$=0.9). 
This is likely due to the simple band structure 
assumed in the above calculation, 
while calculations for Na$_x$CoO$_2$ suggest 
more complicated band structure.\cite{NCO_5}

For the related compounds [Ca$_2$CoO$_3$]$_{0.62}^{\rm RS}$[CoO$_2$] 
and [Ca$_2$Co$_{4/3}$Cu$_{2/3}$O$_4$]$_{0.62}^{\rm RS}$[CoO$_2$], 
({\sl i.e.} 3-layer and 4-layer cobaltites, 
where RS denotes the rocksalt-type susbsystem), 
the {\sf IC-SDW} transition was also observed as a common behavior 
in the CoO$_2$ planes. \cite{muSR_1,muSR_3,muSR_4} 
We have studied by $\mu^+$SR the dependence of the {\sf IC-SDW} transition 
on the Co valence for the variety of layered cobaltites shown in Table~1. 
It should be noted that the magnetic susceptibility $\chi(T)$ curves 
for the 3- and 4-layer samples lacked a marked change 
at either the $T_{\sf SDW}^{\rm on}$ or $T_{\sf SDW}^{\rm end}$ 
detected by $\mu^+$SR. 

\begin{table*}
\caption{\label{tab:table1}Parameters of the magnetic transition 
and thermopower $S$ of several cobaltites; 
the number of the layers between the two adjacent [CoO$_2$] planes ($N$), 
the onset and endpoint temperature of the magnetic transition detected 
by the $\mu^+$SR experiments and $S$ at 300~K. 
nd$\geq$3.3 means not detected down to 3.3 K.
* (**) indicates the sample showing a clear muon precession 
due to the {\sf IC-SDW} ({\sf C-SDW}) field 
in a ZF-$\mu^+$SR spectrum.
}
\begin{ruledtabular}
\begin{tabular}{lcccc}
cobaltite & $N$ & $T_{\sf SDW}^{\rm on}$ (K) & $T_{\sf SDW}^{\rm end}$ (K) & $S$(300~K) ($\mu$VK$^{-1}$)\\
\hline\
[Ca$_2$CoO$_3$]$_{0.62}^{\rm RS}$[CoO$_2$]
&3&100$\pm$5&30$\pm5^{\ast}$&130\\\
[Ca$_{1.8}$Bi$_{0.2}$CoO$_3$]$_x^{\rm RS}$[CoO$_2$]
&3&120$\pm$20&70$\pm10^{\ast}$&140\\\
[Ca$_{1.8}$Y$_{0.2}$CoO$_3$]$_x^{\rm RS}$[CoO$_2$]
&3&120$\pm$20&70$\pm10^{\ast}$&140\\\
[Ca$_{1.8}$Sr$_{0.2}$CoO$_3$]$_x^{\rm RS}$[CoO$_2$]
&3&100$\pm$10&45$\pm10^{\ast}$&118\\\
[Pb$_{0.4}$Co$_{0.6}$Ca$_2$O$_3$]$_{0.62}^{\rm RS}$[CoO$_2$]
&3&130$\pm$20&40$\pm10^{\ast}$&165\\\
[Pb$_{0.7}$Co$_{0.4}$Sr$_1$Ca$_{0.9}$O$_3$]$_{0.58}^{\rm RS}$[CoO$_2$]
&3&90$\pm$10&10$\pm$10&130\\\
[Pb$_{0.7}$Co$_{0.3}$Sr$_2$O$_3$]$_{0.56}^{\rm RS}$[CoO$_2$]
&3&80$\pm$10&nd$\geq$3.3&120\\\
[Tl$_{0.76}$Co$_{0.33}$Sr$_2$O$_3$]$_{0.56}^{\rm RS}$[CoO$_2$]
&3&10$\pm$40&nd$\geq$5.1&90\\\
[Tl$_{1.01}$Co$_{0.11}$Sr$_{1.88}$O$_3$]$_{0.56}^{\rm RS}$[CoO$_2$]
&3&25$\pm$5&nd$\geq$1.7&90\\\
[Tl$_{1.18}$Co$_{0.03}$Sr$_{1.78}$O$_3$]$_{0.57}^{\rm RS}$[CoO$_2$]
&3&nd$\geq$2.5&nd$\geq$2.5&90\\\
[Hg$_{0.4}$Co$_{0.6}$Sr$_2$O$_3$]$_{0.56}^{\rm RS}$[CoO$_2$]
&3&40$\pm$25&5$\pm$5&100\\\
[Ti$_{0.4}$Co$_{0.6}$Ca$_2$O$_3$]$_{0.62}^{\rm RS}$[CoO$_2$]
&3&150$\pm$10&15$\pm$5&150\\
\hline\
[Ca$_2$Co$_{4/3}$Cu$_{2/3}$O$_4$]$_{0.62}^{\rm RS}$[CoO$_2$]
&4&190$\pm$10&145$\pm10^{\ast}$&150\\\
[Ca$_2$Bi$_{1.7}$Co$_{0.3}$O$_4$]$_{0.60}^{\rm RS}$[CoO$_2$]
&4&60$\pm$20&nd$\geq$5&140\\\ 
[Sr$_2$Bi$_2$O$_4$]$_x^{\rm RS}$[CoO$_2$]
&4&nd$\geq$2.3&nd$\geq$2.3&134\\\
[Ba$_2$Bi$_2$O$_4$]$_{0.5}^{\rm RS}$[CoO$_2$]
&4&nd$\geq$1.8&nd$\geq$1.8&100\\ 
\hline\
Na$_{0.9}$CoO$_{2}$&1&19$\pm$0.5&19$\pm3^{\ast}$&150\\\
Na$_{0.75}$CoO$_{2}$&1&22$\pm$0.5&22$\pm1^{\ast\ast}$
&120\\\
Na$_{0.65}$CoO$_{2}$&1&nd$\geq$2.5&nd$\geq$2.5&100\\\
Na$_{0.6}$CoO$_{2}$&1&nd$\geq$4.3&nd$\geq$4.3&100\\
\end{tabular}
\end{ruledtabular}
\end{table*}
There are two Co sites in these layered cobaltite lattices, 
one in the rocksalt-type subsystem and the other in the CoO$_2$ plane. 
Therefore it is difficult to determine the Co valence 
in the CoO$_2$ plane by $\chi$ measurements or 
chemical titration techniques. 
Since the Hall coefficients $R_{\rm H}$ of these cobaltites are 
reported to depend strongly on temperature,\cite{BSCO_1,NCO_4} 
it is also difficult to estimate the carrier concentration 
in the CoO$_2$ plane. 
On the other hand, the magnitude of $S$ in the layered cobaltites 
is almost independent of $T$ above 200~K.
\cite{CCO_2, PbCSCO_1,TlSCO_1,HgSCCO_1,BSCO_1,CCCO_1} 
Therefore, we use the value of $S$ at 300~K 
as an indicator of the Co valence in the CoO$_2$ plane 
for these cobaltites. 
The modified Heikes formula gives $S$ 
at the high-temperature limit ($S_{T\rightarrow\infty}$) as \cite{S_1}
\begin{eqnarray}
 S_{T\rightarrow\infty} &=&
 -\frac{k_{\rm B}}{e} {\rm ln}\biggl(\frac{g_3}{g_4}\frac{y}{1-y}\biggr) ,
\label{eq:tep}
\end{eqnarray}
where $k_{\rm B}$ is the Boltzmann constant, 
$e$ is the elementary charge and 
$g_3$ and $g_4$ are the numbers of the spin configurations 
of Co$^{3+}$ and Co$^{4+}$ ions, respectively, and $y$ is the ratio 
Co$^{4+}$/(Co$^{3+}$+Co$^{4+}$). 
Since both Co$^{3+}$ and Co$^{4+}$ are in the low-spin state 
($t_{2g}^6$ and $t_{2g}^5$) and $g_3$=1 and $g_4$=6, 
we can convert $S$(300~K) to $y$, 
assuming that $S$(300~K)=$S_{T\rightarrow\infty}$. 

\begin{figure}
\includegraphics[width=8cm]{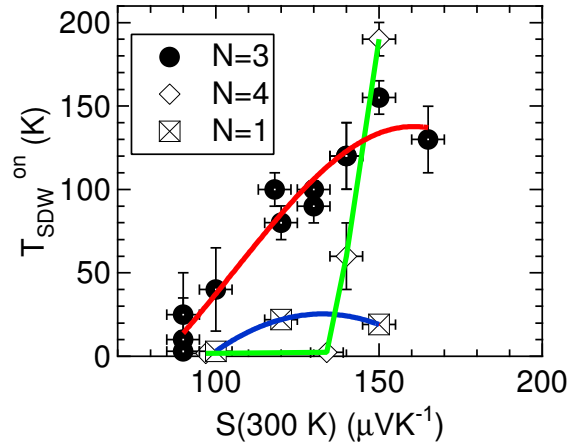}
\caption{\label{fig:ZF-muSR}  The relationship between 
  $T_{\sf SDW}^{\rm on}$ and thermopower $S$ at 300~K. 
  Solid circles represent the data for the cobaltites 
  with a triple rocksalt-type subsystem, 
  open diamonds a quadruple rocksalt-type subsystem 
  and crossed squares Na$_x$CoO$_2$.
}
\end{figure}
Figure~4 shows $T_{\sf SDW}^{\rm on}$ as a function of 
$S$(300~K) for the all cobaltites listed in Table~1.
A clear dome-shaped relation is thus observed 
for all the available cobaltites with a variable number of layers ($N$) 
between the two adjacent CoO$_2$ planes. 
As $N$ increases from 1, the $T_{\sf SDW}^{\rm on}-vs.-S$(300~K) curve 
shifts towards higher temperature. 
This is due to the increased two-dimensionality induced 
by the increase in the interlayer distance between CoO$_2$ planes. 
Also, the large observed transition widths (50-130~K) are 
consistent with enhanced two-dimensionality 
and resulting spin fluctuations. 
Phenomenologically, the phase diagram is very similar to 
the well-known relationship between the superconducting $T_{\rm c}$ and 
the Cu valence in the high-$T_{\rm c}$ cuprates. 
Actually, both {\sf SDW} and superconducting transitions are 
induced by an intrinsic instability of an electron system; 
in other words, as $T$ decreases, 
an energy gap appears at $T_{\sf SDW}$ and/or $T_{\rm c}$ 
to minimize the internal energy for both cases. 
Therefore, it is reasonable to expect a similar relationship 
between transition temperature and carrier concentration 
for both the magnetic cobaltites and the superconducting cuprates.  

Furthermore, the average Co valence for the maximum $T_{\sf SDW}$ 
indicates the optimal filling to induce an {\sf SDW} transition 
at high temperatures and enhance the effective mass of charge carriers 
through the {\sf AF} interaction between spins. 
In other words, this dome relation provides important guidance 
in the search for improved thermoelectric properties 
of the layered cobaltites. 
 
We thank S.R. Kreitzman, B. Hitti, D.J. Arseneau, 
Y. Seno, K. Dohmae, C. Xia, H. Nozaki, H. Hazama, 
J. Chakhalian, D. Liu, A. I.-Najafabadi and S. D. LaRoy 
for help with the $\mu^+$SR experiments. 
Also, we appreciate R. Asahi, U. Mizutani, H. Ikuta, 
T. Takeuchi and K. Machida for discussions. 
This work was supported at Toyota CRDL 
by joint research and development with 
International Center for Environmental Technology Transfer 
in 2002-2004, commissioned 
by the Ministry of Economy Trade and Industry of Japan, 
at UBC by the Canadian Institute for Advanced Research, 
the Natural Sciences and Engineering Research Council of Canada, 
and at TRIUMF by the National Research Council of Canada.


\end{document}